\documentclass[aps,prd,twocolumn,floatfix,nofootinbib,showpacs,superscriptaddress]{revtex4-1}
\usepackage{amsmath,amsfonts,amssymb,bm}
\usepackage{graphicx}
\usepackage{subfigure}
\usepackage{color}
\definecolor{purple}{rgb}{0.5,0,0.5}
\definecolor{blue}{rgb}{0.0,0,0.9}
\usepackage[colorlinks=true, pdfstartview=FitV, linkcolor=purple, citecolor= purple, urlcolor=blue]{hyperref}

\begin{document}
\title{Interface Effect in QCD Phase Transitions via Dyson-Schwinger Equation Approach}

\author{Fei Gao }
\affiliation{Department of Physics and State Key Laboratory of Nuclear Physics and Technology,
Peking University, Beijing 100871, China}
\affiliation{Collaborative Innovation Center of Quantum Matter, Beijing 100871, China}

\author{Yu-xin Liu }
\affiliation{Department of Physics and State Key Laboratory of Nuclear Physics and Technology,
Peking University, Beijing 100871, China}
\affiliation{Collaborative Innovation Center of Quantum Matter, Beijing 100871, China}
\affiliation{Center for High Energy Physics, Peking University, Beijing 100871, China}

\date{\today}

\begin{abstract}
With the chiral susceptibility criterion we obtain the phase diagram of strong-interaction matter in terms of temperature and chemical potential in the framework of Dyson-Schwinger equations (DSEs) of QCD.
After calculating the pressure and some other thermodynamic properties of the matter in the DSE method, we get the phase diagram in terms of temperature and baryon number density.
We also obtain the interface tension and the interface entropy density to describe the inhomogeneity of the two phases in the coexistence region of the first order phase transition.
After including the interface effect, we find that the total entropy density of the system increases
in both the deconfinement (dynamical chiral symmetry restoration) and the hadronization (dynamical chiral symmetry breaking) processes of the first order phase transitions
and thus solve the entropy puzzle in the hadronization process.
%
\end{abstract}



\maketitle

\section{Introduction}

The phase transitions of strong-interaction matter (QCD phase transitions) with respect to temperature $T$ and chemical potential $\mu$ have been investigated for a long time (see, {\it e.g.}, Refs.~\cite{Pisarski:1984PRD,Rajagopal:1999NPA,DOE2008frontiers,Wambach:2009RMP,Fukushima:2011RPP,Owe:2013PPNP}). On theoretical side, the studies include  effective model calculations (see, {\it e.g.}, Refs.~\cite{Buballa:2005PR,Ratti:2006PRD,Schaefer:2007PRD,Fu:2008PRD,Ciminale:2008PRD,Fukushima:2008PRD,Pisarski:1984PRD, Abuki:2008PRD,Schaefer:2009PRD,Costa:2008PRD,Sasaki:2008PRD,Hatta:2003PRD,Zhao:2008EPJC,Mao:2010JPG,Ayala:2011PRD,Jiang:2013PRD,Xin:2014PRDa}), the Dyson-Schwinger equation method (see, {\it e.g.}, Refs.~\cite{Bender:1996PRL,Bender:1998PLB,Blaschke:1998PLB,Maris:2003EPJA,Chen:2008PRD,Fischer:2009PRLa,Qin:2011PRL,Xin:2014PRDb,Zong:2014FBS,Zong:20156PRD,Fischer:201134PLB,Fischer:2014NPA,Mueller:2010EPJC,Qin:2011PRD,Qin:2013PRD,Gao:2014PRD,Bashir:2014JPG,Gao:2016PRD,Fischer:2016PRD,Gao:2016ar}), functional renormalization group approach (see, {\it e.g.}, Refs.~\cite{Pawlowski:2007AP,Pawlowski:2013PRD,Pawlowski:2015PRDa,Pawlowski:2015PRDb,Pawlowski:2015PRL,Pawlowski:20156PRD})  and lattice QCD simulations (see, {\it e.g.}, Refs.~\cite{Karsch:1996NPB,Karsch:2002LNP,Fodor:20024JHEP,DElia:2003PRD,Bernard:2005PRD,Aoki:2006Nature,Aoki:2006PLB,Allton:2006PRD,Forcrand:20023NPB,Endrodi:2011JHEP,Kaczmarek:2011PRD,deForcrand:2014PRL,Karsch:2011PLB,Cea:2014PRD,Ding:2014PRL,Gavai:2005PRD,Li:2011PRD,Bazavov:2012PRD,Gupta:2014PRD,Cheng:2009PRD,Gavai:2011PLB,Bazavov:2012PRL,Borsanyi:2012JHEP,Borsanyi:2013PRL,Borsanyi:2014PRL}).
Based on these researches, it has been widely accepted that at  physical quark mass, the chiral phase transition is a crossover at low chemical potential, while it is a first order phase transition at high chemical potential.
Meanwhile, the confinement--deconfinement phase transition coincides with the chiral phase transition (see, {\it e.g.}, Refs.~\cite{Fischer:201134PLB,Xin:2014PRDa,Gao:2016PRD,Gao:2016ar}).

The first order phase transition is generally described by process of either nucleation or explosive spinodal decomposition~\cite{Palhares:2010PRD}. With the chiral susceptibility criteria (see, {\it e.g.}, Refs.~\cite{Holl:1999PRC,Zhao:2008EPJC,Fischer:2009PRLa,Qin:2011PRL,Ayala:2011PRD}), one usually find the phase transition as a nucleation process which is defined as the transition from a metastable phase to a stable phase, and  the stable phase boundary and the metastable phase boundary determine the coexistence region.
Astrophysical observables of compact stars also favor the nucleation of quark matter~\cite{Heiselberg:1993PRL,Palhares:2011jd,Palhares:2010PRD,Fraga:2015PRD,Weber:2015PRC,Weber:2016PRC}. When the first order phase transition takes place, the two phases with different thermal properties,  the dynamical chiral symmetry breaking (DCSB or Nambu) phase and the dynamical chiral symmetry preserved (DCS or Wigner) phase,  meet at an interface. 
This interface effect, which  is measured by the interface tension and related quantities such as the interface entropy and  the critical size of the bubble~\cite{Ke:2014PRD,Randrup:2009PRC}, has been investigated by lattice QCD simulation~\cite{deForcrand:2004jt} and many effective model calculations~\cite{Palhares:2011jd,Palhares:2010PRD,Ke:2014PRD, Randrup:2009PRC,Pinto:2012PRC,Heiselberg:1993PRL,Mintz:2013PRD,Song:2010PRC,Peng:2010PRC,Lugones:2013PRC,Pinto:2013PRC,Fraga:2015PRD,Weber:2015PRC,Peng:2016PRD,Weber:2016PRC}.
On the other hand, it has been found for a long time that  in the hadronization process there exists a  so called entropy puzzle, that is, the entropy density of the quark-gluon phase is always larger than that of the hadron phase in both the hadronization (DCS to DCSB) process and the deconfinement (DCS restoration) process~\cite{Blaizot:2001PRD,Greco:2003PRC,Yamazaki:2015NPA,Ke:2014PRD,Song:2010PRC}, and thus, the hadronization process seems to be impossible according to the increasing entropy principle. Effective model calculations have provided hints for that considering the interface entropy might solve this puzzle~\cite{Ke:2014PRD,Song:2010PRC}.
To make this solid, it is imperative to study the problem and solve the puzzle via sophisticated continuum QCD approach.

It has been known that the Dyson-Schwinger equations (DSEs), a nonperturbative method of QCD~\cite{Roberts:DSEinitial,Alkofer:2001Infrared,Roberts:20124Review},
are successful in describing  QCD phase transitions ({\it e.g.}, Refs.~\cite{Bender:1996PRL,Bender:1998PLB,Blaschke:1998PLB,Maris:2003EPJA,Chen:2008PRD,Fischer:2009PRLa,Qin:2011PRL,Xin:2014PRDb,Zong:2014FBS,Zong:20156PRD,Fischer:201134PLB,Fischer:2014NPA,Mueller:2010EPJC,Qin:2011PRD,Qin:2013PRD,Gao:2014PRD,Bashir:2014JPG,Gao:2016PRD,Fischer:2016PRD,Gao:2016ar,Roberts:20124Review,Wang:2012PRD}) and hadron properties (for reviews, see Ref.~\cite{Roberts:20124Review}).
We then, in this paper, take the DSE method to calculate the uniform entropy density directly
and include the interface entropy in the free energy expression approximation with the particle number density determined by the DSE method being the input.
%
%
We find that  the interface entropy induced by the interface production is significant to solve the entropy puzzle. As the interface part is taken into account, the total entropy density of the two phases switches into the right order which drives the quark-gluon phase into the hadron phase during the hadronization process.
We also find that the interface entropy is proportional to the area, and this leads to a natural explanation to the area law of the entropy\cite{Srednicki:1993PRL,Eisert:2010RMP}.

The remainder of this paper is organized as follows.  In Section~\ref{DSEs} we describe briefly the framework of DSEs of QCD at finite temperature $T$ and finite chemical potential $\mu$.
 In Section~\ref{phase} we display the results on the phase transitions and the thermodynamic properties. Then we introduce the interface thermodynamics and depict the results in Section~\ref{interface}.
Finally, Section~\ref{sum} provides a summary and remark.

\section{Quark Gap Equation and Thermodynamic Property}
\label{DSEs}

In the framework of DSEs, the quark gap equation at finite temperature and quark chemical potential reads
\begin{eqnarray}
\label{eq:gap1}
S(\vec{p},\tilde\omega_n)^{-1} &=&  i\vec{\gamma}\cdot\vec{p}
+ i\gamma_4 \tilde \omega_{n} + m_{0}^{}
  + \Sigma(\vec{p}, \tilde\omega_{n}) \, ,\\
\nonumber
\Sigma(\vec{p},\tilde\omega_n) &=& T\sum_{l=-\infty}^\infty \! \int\frac{d^3{q}}{(2\pi)^3}\; {g^{2}} D_{\mu\nu} (\vec{p}-\vec{q}, \Omega_{nl}; T, \mu)   \quad \\
& & \times \frac{\lambda^a}{2} {\gamma_{\mu}} S(\vec{q},
\tilde\omega_{l}) \frac{\lambda^a}{2}
\Gamma_{\nu} (\vec{q}, \tilde\omega_{l},\vec{p},\tilde\omega_{n})\, ,
\label{eq:gap2}
\end{eqnarray}
where $m_{0}^{}$ is the current quark mass, $\tilde\omega_{n}^{} = \omega_{n}^{} + i \mu$ with $\omega_{n}^{}=(2n+1)\pi T$ being the quark Matsubara frequency, $\mu$ the quark chemical potential, and $\Omega_{nl} = \omega_{n} - \omega_{l}$. $D_{\mu\nu}$ is the dressed-gluon propagator, and $\Gamma_{\nu}$ is the dressed quark-gluon interaction vertex.
%
%

The gap equation's solution can be decomposed as
\begin{eqnarray}\label{eq:qdirac}
\nonumber
S(\vec{p},\tilde\omega_{n})^{-1} & = & i\vec{\gamma} \cdot \vec{p}\, A(\vec{p}\,^2, \tilde\omega_{n}^2) \\
&& + i\gamma_{4} \tilde\omega_{n} C(\vec{p}\,^2, \tilde\omega_{n}^2) + B(\vec{p}\,^2, \tilde\omega_{n}^2) \, .
\end{eqnarray}
%

The dressed-gluon propagator has the form
\begin{equation}
g^2 D_{\mu\nu}(\vec{k}, \Omega_{nl}) = P_{\mu\nu}^{T} D_{T}(\vec{k}\,^2, \Omega_{nl}^2) + P_{\mu\nu}^{L} D_{L}(\vec{k}\,^2, \Omega_{nl}^2)\,,
\end{equation}
where $P_{\mu\nu}^{T,L}$ are, respectively, the transverse and longitudinal projection operators, and
\begin{equation}
{D_{T}(k_{\Omega})} =\mathcal{D}({k^{2}_{\Omega}},0), \qquad
{D_{L}(k_{\Omega})} =\mathcal{D}({k^{2}_{\Omega}},{m_{g}^{2}}) \, , \quad
\end{equation}
where  $m_{g}^{}$ is the thermal mass of the gluon and can be taken as $m_{g}^{2}=16/5(T^2+6\mu^2/(5\pi^2))$ according to  perturbative QCD calculations~\cite{Haque:2013PRD,Thoma:1998NPA}.

We employ the infrared constant model (Qin-Chang model)~\cite{Qin:2011PRC} for the dressed-gluon propagator that coincides with the results of modern DSEs calculations and lattice QCD sinulations\cite{Bowman:2004PRD,Bogolubsky:2009PLB,Boucaud:2010PRD,Oliveira:2011JPG,Cucchieri:2012PRD,Aguilar:2012PRD,Ayala:2012PRD,Dudal:2012PRD,Strauss:2012PRL,Weber:2012JPCS,Zwanziger:2013PRD,Blossier:2013PRD}, which reads:
\begin{eqnarray}
\mathcal{D}({k^{2}_{\Omega}}, {m_{g}^{2}}) & = & 8{\pi^{2}} D
\frac{1}{\omega^{4}} e^{-{s_{\Omega}^{}}/\omega^{2}}  \nonumber  \\
& & + \frac{8{\pi^{2}} {\gamma_{m}}}{{\ln}[ \tau \! + \! (1 \! + \!
{s_{\Omega}^{}}/{\Lambda_{\text{QCD}}^{2}} ) ^{2} ] } \,
{\cal F}(s_{\Omega}^{}) \, ,
\end{eqnarray}
with ${\cal F}(s_{\Omega}) = (1-\exp(-s_{\Omega}/4 m_{t}^{2}))/s_{\Omega}$, $s_{\Omega}^{} = \Omega^2 + \vec{k}\,^2 + m^{2}_{g}$, $\tau=e^2-1$, $m_t=0.5\,$GeV, $\gamma_m=12/25$, and $\Lambda^{}_{\text{QCD}}=0.234\,$GeV.

To include the temperature screening effect which would screen the interaction when we calculate the thermodynamic properties of QCD, we remedy the coupling $D$ to $D(T,\mu)$ in the similar way as that in Refs.~\cite{Qin:2011PRD,Gao:2014PRD},
\begin{eqnarray}
D(T,\mu)=\left\{
\begin{array}{ll}
\displaystyle
D \,, &   T<T_{\text{p}} \, ,  \\
\displaystyle
\frac{a}{b(\mu)+ \ln[T^\prime/\Lambda_{QCD}]}\,, &  T \ge
T_{\text{p}} \, ,
\end{array}
\right.  \label{DTfunction}
\end{eqnarray}
where  $T^{\prime}=\sqrt{T^2+6\mu^2/(5\pi^2)}$, $T_{\rm p}$ is  the temperature that the thermal screening effect emerges.  At $\mu=0$ we take $T_{\rm p}=T_{c}(\mu=0)$ with $T_c$ the chiral phase transition temperature and $a=0.029$, $b=0.432$, while the phase transition temperature $T_c(\mu)$ would change as chemical potential changes, herein we still apply $T_{\rm p}=T_{c}(\mu)$ and adjust the value of $b$ at every chemical potential to make the coupling strength $D(T_{c}(\mu),\mu)=D$.

For the quark-gluon interaction vertex, we take the approximation $\Gamma_\nu(\vec{q}, \tilde\omega_{l},\vec{p},\tilde\omega_{n}) = \gamma_\nu\,$, which has been widely implemented in Dyson-Schwinger equation calculations and shown to be a quite good approximation in studying hadron properties~\cite{Munczek:1995PRD,Bender:1996PLB,Cloet:2007pi,Burden:1997PRC,Watson:2004FBS,Maris:2007AIPCP,Fischer:2009PRLb,Krassnigg:2009PRD,Krassnigg:2011PRD,Krassnigg:20115PRD,Krassnigg:2015PRD,Krassnigg:20156PRD} and QCD phase transitions (for a comparison with the result in a sophisticated vertex, see, {\it e.g.}, Ref.~\cite{Gao:2016ar}).

The quark pressure is given by the tree-level auxiliary-field
effective action \cite{Haymaker:1990vm} in stationary phase approximation,
\begin{equation}
\label{eq:pressure}
P[S]=\frac{T}{V} \ln{Z}=\frac{T}{V} \Big( Tr\ln{[\beta S^{-1}]}-\frac{1}{2}Tr[\Sigma S] \Big)\, ,
\end{equation}
where $\beta=1/T$ and the self energy $\Sigma=S^{-1}-S^{-1}_0$ with $S_0$ the free quark propagator. However, this definition holds ultraviolet divergence which should be subtracted to get the physical pressure.

Herein we take the subtraction scheme according to the relation~\cite{kapusta1993finite}
\begin{eqnarray}
\label{eq:continuty}
&& T \sum^{\infty}_{n=-\infty} f({p_{0}^{}}=i\omega_{n}^{} +\mu) =\int^{i\infty}_{-i\infty} \frac{d{p_{0}^{}}}{2\pi i} f({p_{0}^{}})+\oint  \frac{d{p_{0}^{}}}{2\pi i} f({p_{0}^{}})\notag \\
&& \qquad -\int^{i\infty+\mu+\epsilon}_{-i\infty+\mu+\epsilon}  \frac{d{p_{0}^{}}}{2\pi i}  f({p_{0}^{}}) \frac{1}{e^{ ({p_{0}^{}} - \mu)/T}+1}\notag\\
&& \qquad -\int^{i\infty+\mu-\epsilon}_{-i\infty+\mu-\epsilon} \frac{d{p_{0}^{}}}{2\pi i} \frac{1}{e^{- ({p_{0}^{}}-\mu)/T}+1} \, .
\end{eqnarray}
The second contour integration  on right-hand-side (rhs) in Eq.~(\ref{eq:continuty}) is a term independent of temperature. This term will get a finite value if there is singularity within the contour. For the Nambu solution of the gap equation, owing to the dynamical mass,
there's no singularity within the contour, and thus it contributes only in the Wigner phase.
The last two terms represent the temperature effect, and it can be seen that the exponential damping factor ensures that there's no additional divergence when including the temperature effect.
The first integration term on the rhs in Eq.~(\ref{eq:continuty}) contains the divergence of  the effective action, and hence the thermal properties could be obtained by subtracting this integration from the numerical data, the left-hand-side (lhs).
The practical algorithm to fulfill the subtraction is straightforward~\cite{Gao:2016PRD}:
solve the quark gap equation at a given $(\mu,T)$-pair for a large number of Matsubara frequencies;
at each $(\mu,T)$, obtain smooth interpolations in $p_{0}^{}$ for the three scalar function
in  Eq.~(\ref{eq:qdirac});
and then compute the difference between the sum, lhs and the first integration, rhs.

\section{Phase Diagram and Uniform Thermal Properties}
\label{phase}

We firstly give the phase diagram in the $\mu$--$T$ plane of the two flavor system with  renormalization-group-invariant current-quark mass $m_{u,d}=6\,$MeV
via the chiral susceptibility criterion with the susceptibility being defined in the framework of DSEs as~\cite{Qin:2011PRL,Gao:2016ar}:
\begin{equation}
\chi_{q}^{}=\frac{\partial B(\vec{p}\,^2=0, \tilde\omega_{n}^2)}{\partial m_{0}^{} }.
\end{equation}
%
%
In the calculation, the parameter(s) taken is, as in many previous works,  $\omega = 0.5\,$GeV with
restriction $D \omega = (0.8 \,\textrm{GeV})^{3}$.

\begin{figure}[htb]
\includegraphics[width=0.42\textwidth]{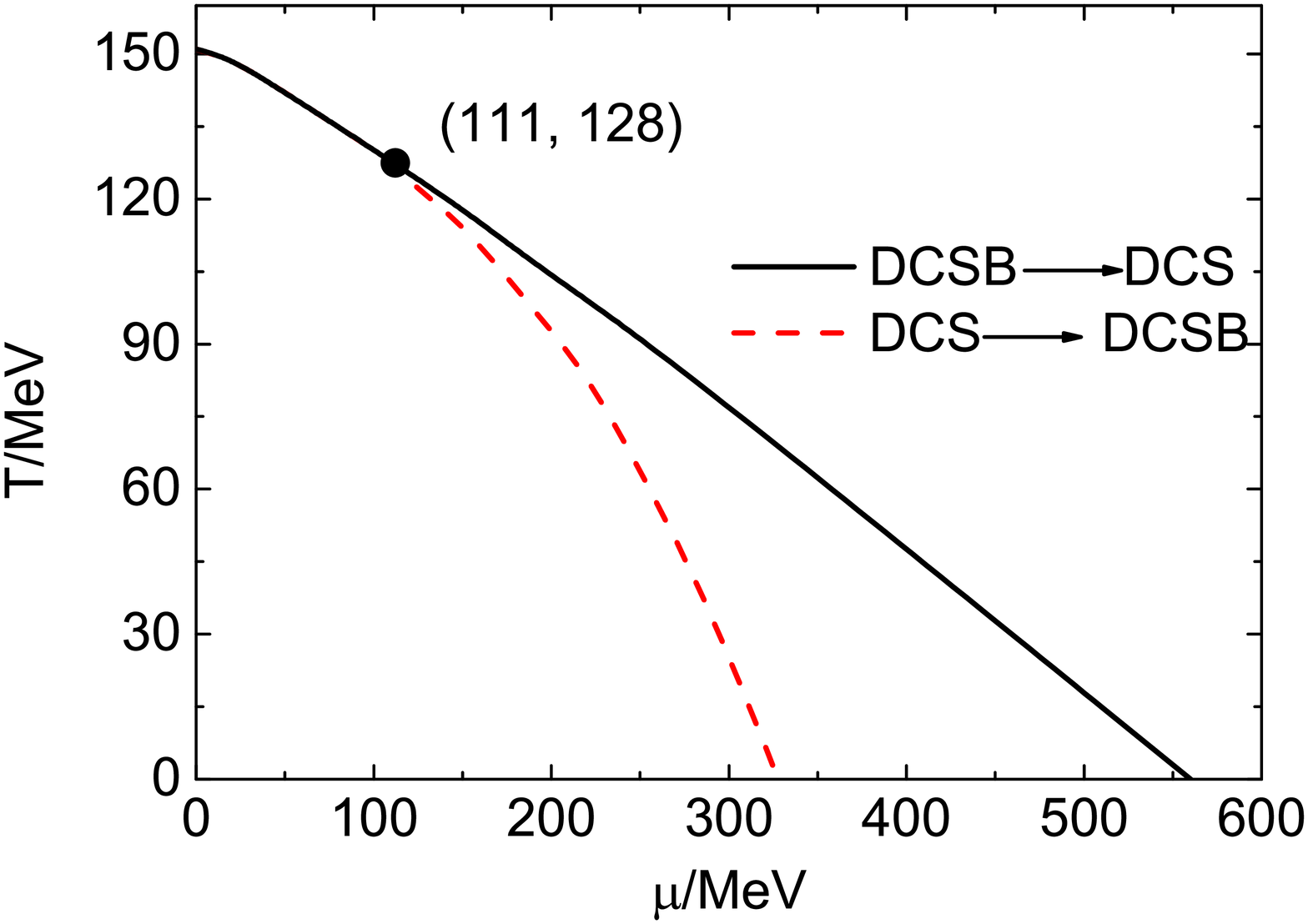}
\caption{(color online) Obtained phase diagram in terms of  temperature and quark chemical potential.}
\label{fig:phase}
\end{figure}

It has been known that, in the first order phase transition region, the susceptibility of the DCSB phase and that of the DCS phase diverge at different locations. The set of the states for the susceptibility of each phase to diverge identify a boundary. The region between the two boundaries is just the coexistence region which consists of a stable phase and a metastable phase (in detail, the DCSB phase changes from stable to metastable and the DCS phase varies from metastable to stable, as the chemical potential increases at a certain temperature).
In the crossover region the susceptibility does not diverge and thus, as usual, we define the location for the susceptibility to take its maximum as the phase transition point.
Since the critical end-point (CEP) connects the two regions and thus will combine both the characters,
that is, the susceptibilities of the two phases diverge at the same location.
The obtained phase diagram is shown in Fig.~\ref{fig:phase}.
It is evident that the crossover takes place in the low chemical potential region.
Beyond the location of the CEP $(\mu^{\chi,E}_{q},T^{\chi,E}) = (111, 128)\;$MeV,
the chiral phase transition of QCD becomes first order transition.
Since the nucleation process describes the phase transition from a metastable phase to a stable phase,  we find that the chiral phase transition takes place at different locations for different processes\cite{Ke:2014PRD}.
The DCSB to DCS phase transition takes place at higher chemical potential till the higher bound of the coexistence region, while for the opposite process, hadronization process accomplishes at low chemical potential till the lower bound of the coexistence region.

The phase diagram could be converted into the plane of temperature and baryon number density $n_{B}^{}$ with the relation:
\begin{equation}
n_{B}^{} =\frac{1}{3} n_{q}^{} = \frac{1}{3}\frac{\partial P}{\partial \mu_{q}} \, ,
\end{equation}
where $\mu_{q}$ refers to the quark chemical potential, $n_{q}^{}$ is the quark number density.

\begin{figure}[htb]
\includegraphics[width=0.45\textwidth]{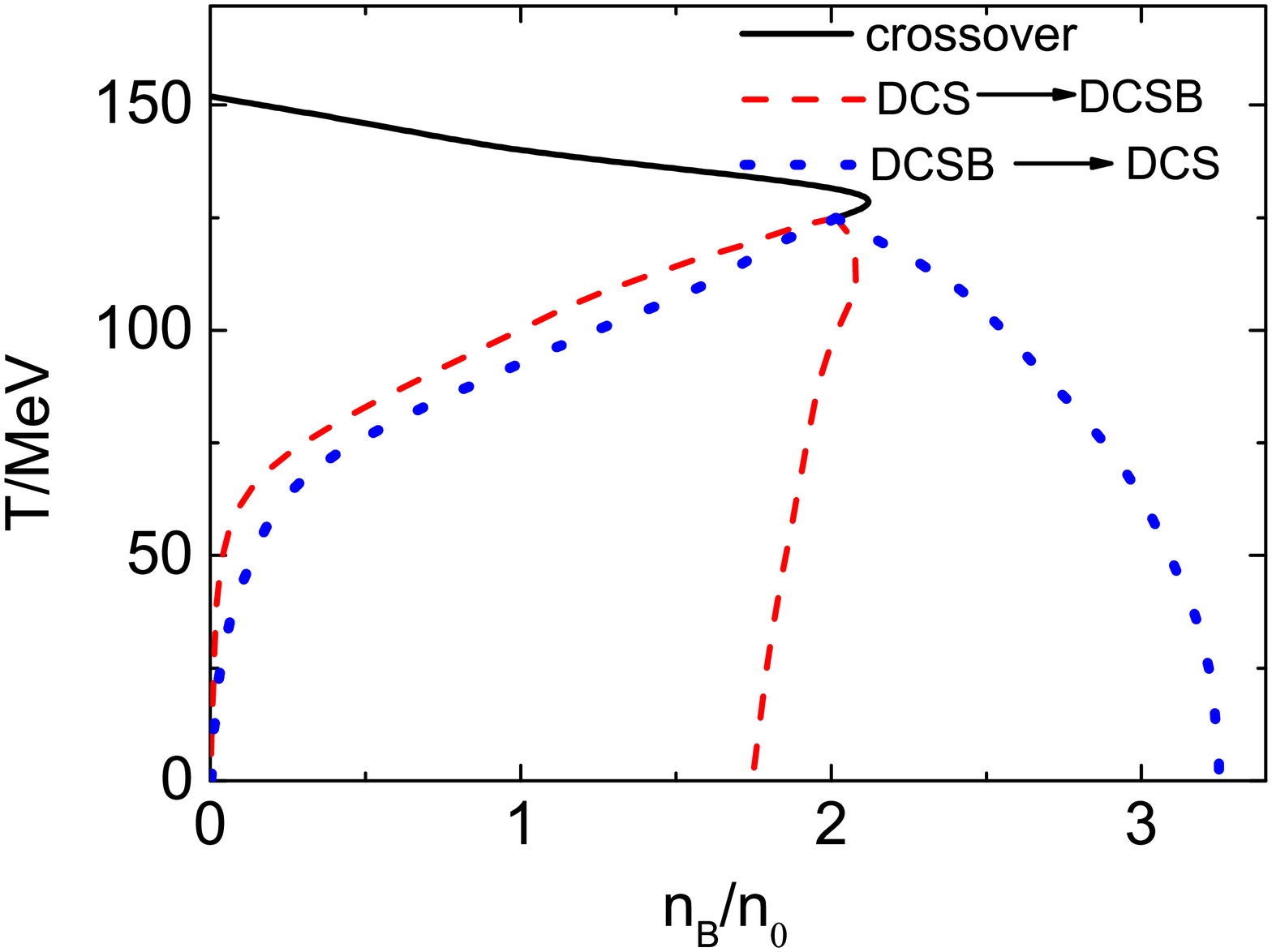}
\caption{(color online) Obtained phase diagram in terms of temperature and baryon number density.}
\label{fig:rhophase}
\end{figure}

The obtained result is displayed in Fig.~\ref{fig:rhophase}.
The number densities of the two phases in the hadronization process are shown as the dashed curve, while those for the opposite process are depicted as the dotted curve.
For each curve, there are two values at every certain temperature corresponding to the different number densities of the two phases.
The larger one is the number density of the DCS phase,
and the smaller one is that of the DCSB phase.
As the temperature increases, they converge gradually at the CEP and become identical
as it becomes crossover.
The energy density at the CEP is $E_{g}=0.377\,\textrm{GeV} \cdot \textrm{fm}^{-3}$
and the baryon number density at CEP is $n_{B}^{}=2.02\, n_{0}^{}$ with $n_{0}^{}=0.16\,$fm$^{-3}$,
the saturation baryon number density of nuclear matter.

After then we obtain the entropy density with the Duhem-Gibbs relation:
\begin{equation}
s_{V}^{}=\frac{1}{T}(\epsilon+P-\mu n)=  \frac{\partial P}{\partial T} .
\end{equation}

\begin{figure}[htb]
\includegraphics[width=0.45\textwidth]{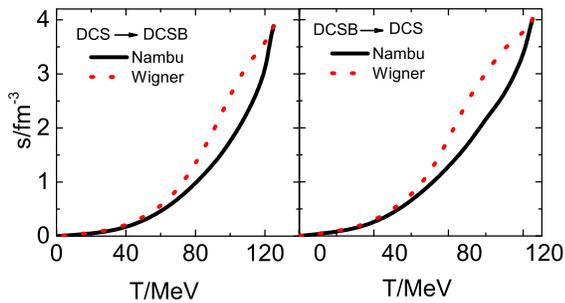}
\caption{(color online) Obtained variation behavior of the $s_{V}^{}$ of the two phases in the first order phase transition region with respective to temperature.
\emph{Left panel} -- in the process from DCS to DCSB phase; and \emph{right panel} -- in the process from DCSB to DCS phase.}
\label{fig:entropy}
\end{figure}

The obtained results are shown in Fig.~\ref{fig:entropy}.
It is evident that, for the phase transition from the DCSB to the DCS phase, the entropy density increases;
while in the opposite process, hadronization process,
the entropy density of the DCS phase is still larger than that of the DCSB phase.
It is apparent that this result violets the increasing entropy principle,
and hence, makes the phase transition from DCS to DCSB (hadronization) unable to take place automatically as temperature decreases,
which is just the so called entropy puzzle in the hadronization process.
People might assume that  after the hadronization process, the released energy will increase the volume of system, and in turn, the total entropy still increases even though the  entropy density of the system decreases.
However, during the phase transition process, when the hadron bubble emerges in the DCS  state, the interface tension will constrain the volume of the bubble to increase.
Thusly, at this moment, the key to solve the entropy puzzle would be the inhomogeneity, {\it i.e.}, the interface between the distinct phases of the system~\cite{Ke:2014PRD}.
When the phase transition takes place, the two phases meet at an interface which gives additional interface entropy.
With this part retrieved for the total entropy, we may find that the total entropy density increases in the hadronization process.

\section{Thermodynamics Including the Interface Effect}
\label{interface}

As mentioned above, with the Duhem--Gibbs relation, we can only get the thermodynamic properties of uniform bulk matters.
Ref.~\cite{Ke:2014PRD} provides a hint that,
to solve the entropy puzzle, it is necessary to consider the effect of the interface.
As the contribution of the interface is taken into account, the total free energy variance of the matter
around the interface holds:
\begin{equation}
\label{eq:freeenergy-general}
dF = -\Delta PdV - S_A dT + \gamma dA,
\end{equation}
where $\Delta P$ is the pressure difference between the two phases,
 $S_{A}^{}$  the interface entropy,
$\gamma$  the interface tension and $A$  the interface area.
For a stable interface, with Maxwell relation we obtain straightforwardly the relation between the interface entropy $S_{A}^{}$ and interface tension as:
\begin{equation}
\label{eq:max}
S_{A}^{} = A\cdot s_{A}^{} = A \left(\frac{\partial S_{A}^{}}{\partial A}\right)_{V,T}
= - A\left(\frac{\partial \gamma}{\partial T}\right)_{V,T} \, ,
\end{equation}
where $s_{A}^{}$ is the interface entropy density.

The total entropy density of the system could then be written as:
\begin{equation}
s_{tot}^{}(T) = s_{V}^{} + \frac{A}{V} s_{A}^{},
\end{equation}
where $V$, $A$ is, respectively, the volume of the system  and the area of the interface.
The ratio $A/V$ is hard to obtain completely and accurately.
However, we can intuitively make an assumption that the system is composed of lattices with length $2R$, and each of the lattice includes a spherical bubble with radius $R$.
Then the relation between the $A$ and the $V$ can be approximately given as
$A/V \approx \pi /(2R)$, where the $R$ is the average radius of the bubbles formed during the phase transition process.
Such a radius can be extracted from the stationary condition of free energy (combining the geometric property and the result from Eq.~(\ref{eq:freeenergy-general})) as:
\begin{equation}
\label{eq:radius}
\frac{3}{\pi }R=\frac{dV}{dA}=\frac{\gamma}{\Delta P}.
\end{equation}

Taking the phenomenological expansion of free energy density for the first order phase transition system~\cite{Randrup:2009PRC}, we have
\begin{equation}
f(\vec{r})=n\mu+\frac{1}{2}C(\nabla n)^2,
\end{equation}
where $C=\frac{a^2}{n_{B}^{2}}E_{g}$. To evaluate the free energy of our interested system in DSE calculation we take $n_{B}^{}$ and $E_{g}$ to be the baryon number density and the energy density at the CEP. For the parameter $a$, the measure of the interface's thickness, we choose it to be $0.33\,$fm
as that in Refs.~\cite{Ke:2014PRD,Peng:2010PRC}.

The density distribution could be obtained through the variation of free energy under the restriction of normalization according to the equilibration condition:
\begin{equation}
0 = \delta\int \Big{\{} n(\vec{r}) {\mu_{T}^{}} [n] + \frac{1}{2}C(\nabla n)^{2} - \mu_{0}^{} n(\vec{r}) \Big{\}}  d\vec{r} \, ,
\end{equation}
where $\mu_{T}^{}[n]$, $\mu_{0}^{}$ is the distribution of chemical potential, uniform chemical potential, respectively.

The equation of motion for the spherical case  could be  written as
\begin{equation}
 \Delta f_{T}^{} +\frac{1}{2} C \Big{(} \frac{\partial n}{\partial r} \Big{)}^{2} = 0 \, ,
\end{equation}
where $\Delta f_{T}^{} = f_{T}^{}(n) - f_{M}^{}(n)$ is the difference between the uniform free energy density defined in Ref.~\cite{Randrup:2009PRC}, and $f_{M}^{}$ is the Maxwell construction of the free energy defined as:
\begin{equation}
f_{M}^{}(n) = f_{T}^{}(n_{L}^{}) + \frac{f_{T}^{}(n_{H}^{}) - f_{T}^{}(n_{L}^{})}{n_{H}^{} - n_{L}^{}} (n - n_{L}^{}) \, ,
\end{equation}
where $n_{H}^{}$ and $n_{L}^{}$ are the densities of the two coexisting phases.
$\Delta f_{T}^{}$ vanishes at the boundary and positive in between,
and thus can be considered as the free energy gained by undergoing a phase mixture.
The interface tension which is the deficit in free energy per unit interface area
could then be expressed as:
\begin{equation}
\label{eq:tension}
\gamma(T) = \int^\infty_{-\infty} \Delta f_{T}^{} dx = \int^{n_{H}^{}}_{n_{L}^{}}
\sqrt{\frac{C}{2} \Delta f_{T}^{}(n)} dn \, .
\end{equation}

The equation of motion has been used to obtain the second relation in Eq.~(\ref{eq:tension}).
This relation is significant since it indicates that the interface tension is independent of the size scale of the system and thus an intensive quantity.
With the relation in Eq.~(\ref{eq:max}), we find that the interface entropy is proportional to the area of the interface.
Generally, entropy is an extensive quantity with $S\sim R^3$, but the interface entropy satisfies $S_{A}^{} \sim R^{2}$.
Especially in strong coupling limit, the uniform entropy of  strongly interacting particles is small,
and thus the interface part takes upon a great proportion.
The total entropy of the system will then tend to satisfy the area law in the limit.
Therefore, the area law for the entropy indicates that  the system becomes strongly coupled,
and the interface effect  decides most of the thermal properties of the system.
Our present result provides an evidence for the area law of the entropy\cite{Srednicki:1993PRL,Eisert:2010RMP} directly from the view of strong interaction theory (QCD).

\begin{figure}[htb]
\includegraphics[width=0.45\textwidth]{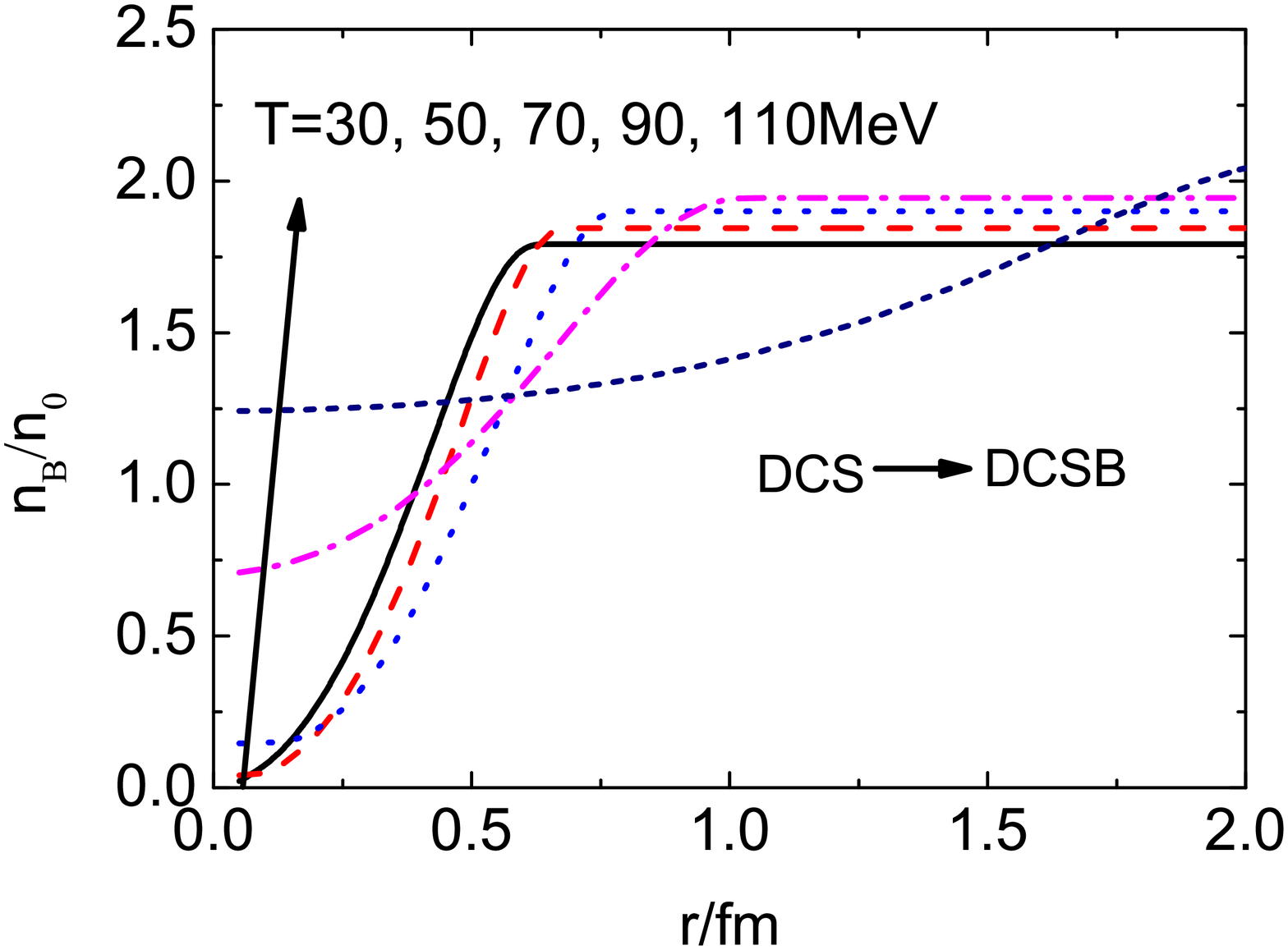}
\includegraphics[width=0.45\textwidth]{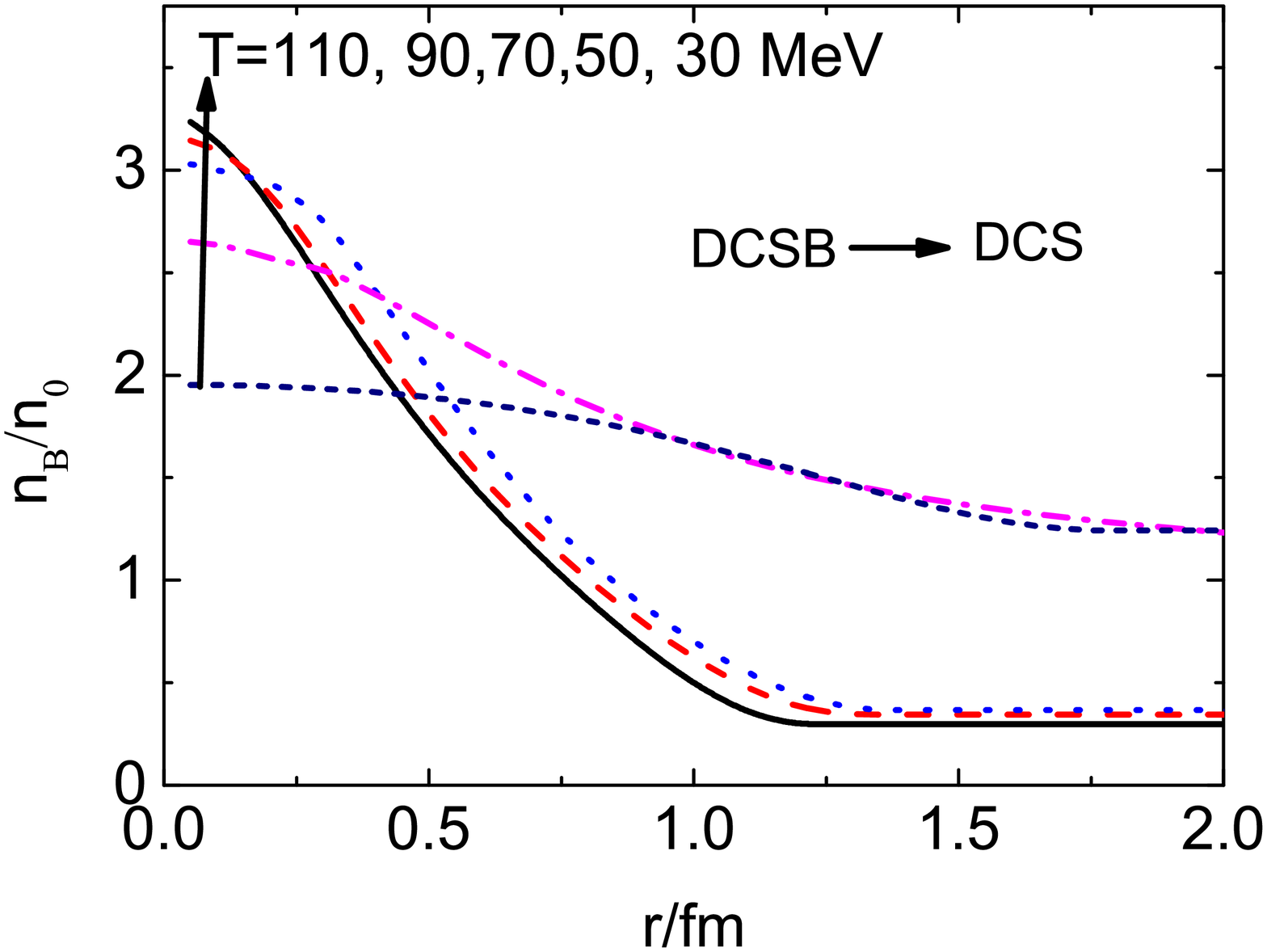}
\caption{(color online) Obtained density distribution at some temperature $T$ in the two processes,
\emph{upper panel} -- for the process from DCS to DCSB phase;
and \emph{lower panel} -- for the process from DCSB to DCS phase.}
\label{fig:density}
\end{figure}

After solving the equation of motion, we get directly the density distribution at every temperature from the DSE approach.
The obtained results of the density distribution are exhibited in Fig.~\ref{fig:density}.
It is apparent that, for the process from the DCS phase to the DCSB phase, the hadronization process, the DCSB phase comes into being inside the bubble, and the DCS phase is outside.
While for the process from DCSB to DCS phase, the density distribution is opposite.

\begin{figure}[htb]
\includegraphics[width=0.48\textwidth]{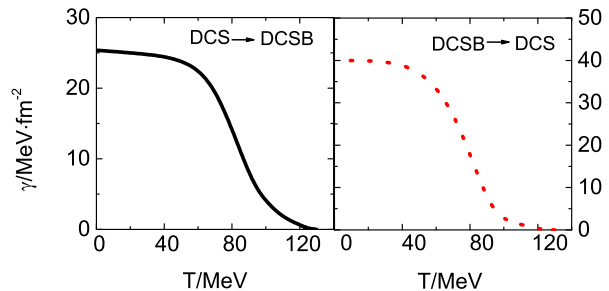}
\caption{(color online) Obtained variation behavior of the interface tension with respective to temperature,
\emph{left panel} -- for the process from DCS to DCSB phase;
and \emph{right panel} -- for the process from DCSB to DCS phase.}
\label{fig:tension}
\end{figure}

The interface tension could be obtained straightforwardly through Eq.~(\ref{eq:tension}).
In Fig.~\ref{fig:tension} we show our results of the temperature dependence of the interface tension for the two processes.
It is evident that the interface tension of hadronization process is smaller than that of the opposite at a certain temperature, due to that the particle density distribution regions are different.
The interface tension at zero temperature is $25.4\, \textrm{MeV}/\textrm{fm}^{2}$
for the hadronization process and $40.0\,\textrm{MeV}/\textrm{fm}^{2}$ for the opposite.
Such results coincide with those given in other calculations ({\it e.g.}, Refs.~\cite{Randrup:2009PRC,Heiselberg:1993PRL,Ke:2014PRD}) excellently.
As the temperature increases, the interface tension decreases monotonically and vanishes near the CEP. To the convenience of being implemented elsewhere, we give approximately the temperature dependence of the  interface tension in the form as:
\begin{equation}
\label{eq:FittedIT}
\gamma(T) = a + b\,e^{(c/T +d/T^2)} \, ,
\end{equation}
with parameters listed in Table.~\ref{tab:fitting}.
\begin{table}[htb]
\centering
\caption{ Fitted parameters  of interface tension in Eq.~(\ref{eq:FittedIT})}\label{tab:fitting}
\begin{tabular}{ccccc}
\hline
 \, & $a/($MeV/fm$^{2})$  & $ b/($MeV/fm$^{2})$ & $c/$MeV & $d/$GeV$^{2}$ \\
\hline
DCS$\rightarrow$DCSB   & $25.4$ & $-1.5$ & $736$ & $-0.048$ \\
DCSB$\rightarrow$DCS   & $40.0$ & $-8.1$ & $399$ & $-0.025$  \\
\hline
\end{tabular}
\end{table}

\begin{figure}[htb]
\includegraphics[width=0.48\textwidth]{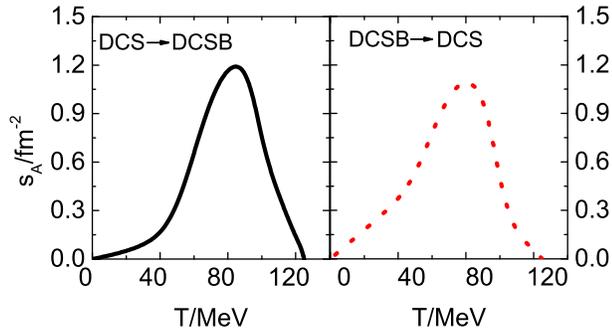}
\caption{(color online) Obtained temperature dependence of the interface entropy density in the two processes.
\emph{Left panel} -- for the process from DCS to DCSB phase;
and \emph{right panel} -- for the process from DCSB to DCS phase.}
\label{fig:interfaceentropy}
\end{figure}

We take then the first order derivative of the interface tension with respect to temperature and obtain the interface entropy density. The obtained results of the temperature dependence of the interface entropy density in the two phase transition processes are shown in Fig.~\ref{fig:interfaceentropy}.
Our above discussion manifests that the interface tension and the interface entropy make contribution in the coexistence region owing to the difference of the particle number densities of the two phases,
thusly it vanishes at CEP.
Fig.~\ref{fig:interfaceentropy} shows obviously that, in both processes, the interface entropy densities experience a rapid increase at intermediate temperature and tend to be zero at zero temperature and at the CEP as expected in general principle.

\begin{figure}[htb]
\includegraphics[width=0.48\textwidth]{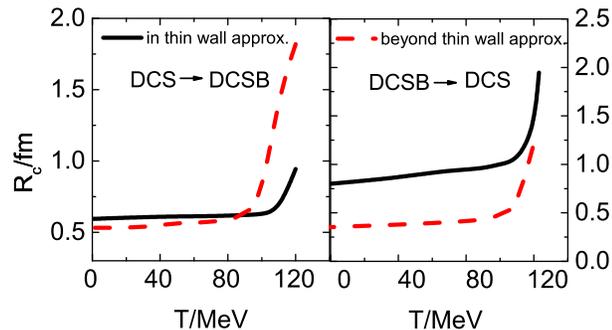}
\caption{(color online) Obtained temperature dependence of the bubble size in the two process,
\emph{left panel} -- in the process from DCS to DCSB phase;
and \emph{right panel} -- in the process from DCSB to DCS phase.}
\label{fig:size}
\end{figure}

After then, we identify the bubble size to obtain the total entropy density.
The typical bubble size during the phase transition could be estimated through Eq.~(\ref{eq:radius}).
We illustrate the obtained results as the solid lines in Fig.~\ref{fig:size}.
It is apparent that, the radius barely changes at low temperature, till about $T\sim 110\,$MeV,
and the radius increases drastically thereafter in both the processes.
In more detail, the radius in hadronization (from DCS to DCSB) process is smaller
at the same temperature since the process takes place at lower chemical potential (density).

\begin{figure}[htb]
\includegraphics[width=0.48\textwidth]{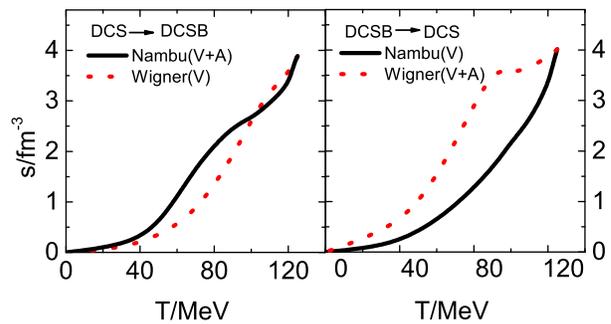}
\caption{(color online) Obtained temperature dependence of the total entropy density of the system with the bubble size determined by Eq.~(\ref{eq:radius}),
\emph{left panel} -- for the process from DCS to DCSB phase;
and \emph{right panel} -- for the process from DCSB to DCS phase.}
\label{fig:entropyt}
\end{figure}

Combining the results obtained above, we can get eventually the contribution of the interface entropy to the total entropy density ($\frac{A}{V}s_{A}^{} = \frac{\pi}{2 R} s_{A}^{}$) of the system.
In Section~\ref{phase} we have got the entropy density of the bulk matter without interface,
which shows that, in both the processes, the entropy density of the DCS phase is larger than that of the DCSB phase,
which violates the increasing entropy principle during the hadronization process.
After adding the contribution of the interface entropy density to the total entropy density,
we find that the total entropy density of the system increases in both the phase transition processes as shown in Fig.~\ref{fig:entropyt}.

\begin{figure}[htb]
\includegraphics[width=0.48\textwidth]{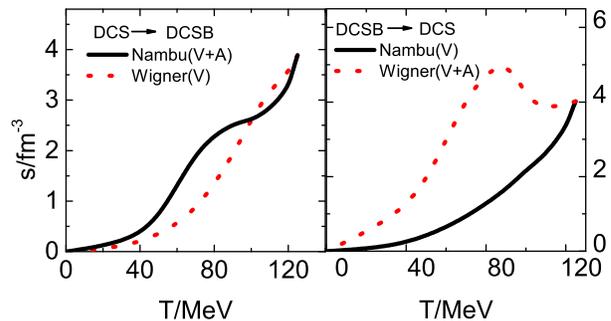}
\caption{(color online) Obtained temperature dependence of the total entropy density with the bubble size determined beyond the thin-wall approximation,
\emph{left panel} -- for the process from DCS to DCSB phase;
and \emph{right panel} -- for the process from DCSB to DCS phase.}
\label{fig:entropyt2}
\end{figure}

Recalling Eq.~(\ref{eq:radius}) and the derivation process, one can notice that we have not yet considered the thickness of the bubbles except for introducing a parameter in the free energy expansion to determine the interface tension. The above result is usually referred to that in thin-wall approximation.
In fact, one can determine the bubble size directly from the density distribution.
That is, namely, that beyond the thin-wall approximation, where the size reads as the radius corresponding to the steepest change of the density distribution (the results at some temperatures have been displayed in Fig.~\ref{fig:density}).
With such a definition of the bubble size, we can also get the radius of the bubble (shown as the dashed line in Fig.~\ref{fig:size}) and the interface entropy density. In turn we obtain the temperature dependence of the total entropy density as illustrated in Fig.~\ref{fig:entropyt2}. It manifests evidently that, when altering the definition of the size, the total entropy density in each of the two phase transition processes changes quantitatively, but not qualitatively.
In both cases, the interface effect modifies the total entropy density of the system and makes the hadronization process coincide with the increasing entropy principle.

\section{Summary}
\label{sum}

In summary, we studied some thermodynamic properties of QCD matter,
%
%
especially those in the first order phase transition region via the Dyson-Schwinger equations approach.
We obtained the phase diagram of the chiral phase transition in terms of the temperature and the chemical potential and that in terms of temperature and baryon number density with a proper subtraction scheme to get the quarks' pressure and the related thermal quantities of the system.
We calculated the entropy densities of both the DCSB and the DCS phases in the first order phase transition region and found that the entropy density of the DCS phase is always larger than that of the DCSB phase in both the phase transition processes of uniform bulk matter.
We then took the free energy expansion scheme with the particle number density distribution obtained in the DSE calculation being the input and include the contribution of the interface demonstrating the inhomogeneity of the coexistence region.
We calculated further the interface tension and the interface entropy density and found an area law for the interface entropy.
After taking the interface effect into account, we observed that the total entropy density increases in both the DSCB to DCS and the DCS to DCSB processes of the first order phase transition
and thus solved the entropy puzzle in the hadronization process.
%
%
In addition, we would like to mention that our present work provides an evidence for
that the violation of the increasing entropy principle in some processes may result from having not taken all the effects, for instance the structure and the entanglement among the ingredients,
completely.

\section{Acknowledgments}
The work was supported by the National Natural Science Foundation of China under Contracts No. 11435001; the National Key Basic Research Program of China under Contract Nos. G2013CB834400
 and 2015CB856900.


\begin{thebibliography}{99}

\bibitem{Pisarski:1984PRD}
     R. D. Pisarski,  F. Wilczek,
        Phys. Rev. D {\bf 29} 338 (1984).

\bibitem{Rajagopal:1999NPA}
     K. Rajagopal,
        Nucl. Phys. A {\bf 661}, 150 (1999).

\bibitem{DOE2008frontiers}
     D. N. S. A. Committee, et al.,
         arXiv:0809.3137 (2008).

\bibitem{Wambach:2009RMP}
     P. Braun-Munzinger, and J. Wambach,
        Rev. Mod. Phys. {\bf 81}, 1031 (2009).

\bibitem{Fukushima:2011RPP}
     K. Fukushima, and T. Hatsuda,
        Rept. Prog. Phys. {\bf 74}, 014001 (2011).

\bibitem{Owe:2013PPNP}
     O. Philipsen,
        Prog. Part. Nucl. Phys. {\bf 70}, 55 (2013).

\bibitem{Hatta:2003PRD}
      Y. Hatta, and T. Ikeda,
          Phys. Rev. D {\bf 67}, 014028 (2003).

\bibitem{Buballa:2005PR}
   M. Buballa,
      Phys. Rept. {\bf 407}, 205 (2005).

\bibitem{Ratti:2006PRD}
     C. Ratti, M. A. Thaler, and W. Weise,
        Phys. Rev. D {\bf 73}, 014019 (2006).

\bibitem{Schaefer:2007PRD}
      B. J. Schaefer, J. M. Pawlowski,  J. Wambach,
          Phys. Rev. D {\bf 76}, 074023 (2007).

 \bibitem{Fu:2008PRD}
     W. J. Fu, Z. Zhang, Y. X. Liu,
        Phys. Rev. D {\bf 77}, 014006 (2008).

\bibitem{Sasaki:2008PRD}
      C. Sasaki,  B. Friman,  K. Redlich,
           Phys. Rev. D {\bf 77}, 034024 (2008).

\bibitem{Ciminale:2008PRD}
     M. Ciminale, R. Gatto, N. D. Ippolito, G. Nardulli, and M. Ruggieri,
       Phys. Rev. D {\bf 77}, 054023 (2008).

\bibitem{Costa:2008PRD}
      P. Costa, M. C.  Ruivo, C. A. de Sousa,
          Phys. Rev. D {\bf 77}, 096001 (2008).

\bibitem{Fukushima:2008PRD}
     K. Fukushima,
        Phys. Rev. D {\bf 77}, 114028 (2008).

\bibitem{Abuki:2008PRD}
     H. Abuki, R. Anglani, R. Gatto, G. Nardulli, and M. Ruggieri,
        Phys. Rev. D {\bf 78}, 034034 (2008).

\bibitem{Zhao:2008EPJC}
     Y. Zhao, L. Chang, W. Yuan, and Y.-x. Liu,
       Eur. Phys. J. C {\bf 56}, 483 (2008).

\bibitem{Schaefer:2009PRD}
     B. J. Schaefer, M. Wagner,
         Phys. Rev. D {\bf 79}, 014018 (2009).

\bibitem{Mao:2010JPG}
    H. Mao, J. S. Jin, and M. Huang,
        J. Phys. G {\bf 37}, 035001 (2010).

\bibitem{Ayala:2011PRD}
     A. Ayala, A. Bashir, C.A. Dominguez, E. Guti\'{e}rrez, M. Loewe, and A. Raya,
        Phys. Rev. D {\bf 84}, 056004 (2011).

\bibitem{Jiang:2013PRD}
    T. Sasaki, Y. Sakai, H. Kouno, and M. Yahiro,
       Phys. Rev. D {\bf 82}, 116004 (2010);
%
    L. J. Jiang, X. Y. Xin, K. L. Wang, S. X. Qin, and Y. X. Liu,
       Phys. Rev. {\bf D 88}, 016008 (2013).
%

\bibitem{Xin:2014PRDa}
     X. Y. Xin, S. X. Qin, and Y. X. Liu,
       Phys. Rev. D {\bf 89}, 094012 (2014).

\bibitem{Bender:1996PRL}
     A. Bender, D.  Blaschke, Y. Kalinovsky, and C. D. Roberts,
       Phys. Rev. Lett. {\bf 77}, 3724 (1996).

\bibitem{Bender:1998PLB}
     A. Bender, G. I. Poulis, C. D.  Roberts, S. M.   Schmidt, and A. W. Thomas,
       Phys. Lett. B {\bf 431}, 263 (1998).

\bibitem{Blaschke:1998PLB}
    D. Blaschke, C. D. Roberts, and S. Schmidt,
       Phys. Lett. B {\bf 425}, 232 (1998).

\bibitem{Maris:2003EPJA}
     P. Maris, A. Raya, C. D. Roberts, and S. M. Schmidt,
       Eur. Phys. J. A {\bf 18}, 231 (2003).

\bibitem{Chen:2008PRD}
     H. Chen, W. Yuan, L. Chang, Y. X. Liu, T. Klahn, and C. D. Roberts,
       Phys. Rev. D {\bf 78}, 116015 (2008).

\bibitem{Fischer:2009PRLa}
     C. S. Fischer,
        Phys. Rev. Lett. {\bf 103}, 052003 (2009);
%
     C. S. Fischer, and J. A. Mueller,
        Phys. Rev. D {\bf 80}, 074029 (2009).

\bibitem{Mueller:2010EPJC}
    J. A. Mueller,  C. S. Fischer,  and D. Nickel,
       Eur. Phys. J. C {\bf 70}, 1037 (2010).

\bibitem{Qin:2011PRL}
   S. X. Qin, L. Chang, H. Chen, Y. X. Liu, and C. D. Roberts,
     Phys. Rev. Lett. {\bf 106}, 172301 (2011).

\bibitem{Qin:2011PRD}
     S. X. Qin, L. Chang, Y. X. Liu, and C. D. Roberts,
       Phys.\ Rev.\ D {\bf 84}, 014017 (2011).

\bibitem{Qin:2013PRD}
     S. X. Qin, and D. H. Rischke,
       Phys. Rev. D {\bf 88}, 056007 (2013).

\bibitem{Gao:2014PRD}
     F. Gao, S. X.  Qin, Y. X. Liu, C. D.Roberts, and S. M. Schmidt,
       Phys. Rev. D {\bf 89}, 076009 (2014).

\bibitem{Xin:2014PRDb}
   X. Y. Xin, S. X. Qin, and Y. X. Liu,
      Phys. Rev. D {\bf 90}, 076006 (2014).

\bibitem{Zong:2014FBS}
     M. He, Y. Jiang, W. M. Sun, and H. S. Zong,
        Phys. Rev. D {\bf 77}, 076008 (2008);
     M. He, F. Hu, W. M. Sun, and H. S. Zong,
        Phys. Lett. B {\bf 675}, 32 (2009);
     B. Wang, Z. F. Cui, W. M. Sun, and H. S. Zong,
        Few-Body Syst. {\bf 55}, 47 (2014);
%
     C. Shi, Y. L. Wang, Y. Jiang, Z. F. Cui, and H. S. Zong,
        J. High Energy Phys. {\bf 1407}, 014 (2014);

\bibitem{Zong:20156PRD}
     S. S. Xu, Z. F. Cui, B. Wang, Y. M. Shi, Y. C. Yang, and H. S. Zong,
        Phys. Rev. D. {\bf 91}, 056003 (2015);
     Y. Lu, Z. F. Cui, Z. Pan, C. H. Chang, and H. S. Zong,
        Phys. Rev. D. {\bf 93}, 074037 (2016).

\bibitem{Fischer:201134PLB}
      C. S. Fischer, J. Luecker, and J. A. Mueller,
          Phys. Lett. B {\bf 702}, 438 (2011).
%
      C. S. Fischer, and J. Luecker,
         Phys. Lett. B {\bf 718}, 1036 (2013).
%
      C. S. Fischer, J. Luecker, and C. A. Welzbacher,
         Phys. Rev. D {\bf 90}, 034022 (2014).

\bibitem{Fischer:2014NPA}
      C. S. Fischer,  J. Luecker,  and C. A. Welzbacher,
         Nucl. Phys. A {\bf 931}, 774 (2014).

\bibitem{Bashir:2014JPG}
    E. Guti\'{e}rrez, A. Ahmad, A. Ayala, A. Bashir, and A. Raya,
        J. Phys. G {\bf 41}, 075002 (2014).

\bibitem{Fischer:2016PRD}
    G. Eichmann, C. S. Fischer, and C. A. Welzbacher,
        Phys. Rev. D {\bf 93}, 034013 (2016).

\bibitem{Gao:2016PRD}
    F. Gao, J. Chen, Y. X. Liu, S. X. Qin, C. D.Roberts, and S. M. Schmidt,
      Phys. Rev. D {\bf 93}, 094019 (2016).

\bibitem{Gao:2016ar}
     F. Gao, and Y. X. Liu,
       arViv:1607.01675.

\bibitem{Pawlowski:2007AP}
      J. M. Pawlowski,
          Ann. Phys. {\bf 322}, 2831 (2007).

\bibitem{Pawlowski:2013PRD}
      L. Fister, and J. M. Pawlowski,
          Phys. Rev. D {\bf 88}, 045010 (2013).

\bibitem{Pawlowski:2015PRDa}
      M. Mitter, J. M. Pawlowski, and N. Strodthoff,
          Phys. Rev. D {\bf 91}, 054035 (2015).


\bibitem{Pawlowski:2015PRDb}
      N. Mueller, and J. M. Pawlowski,
          Phys. Rev. D {\bf 91}, 116010 (2015).

\bibitem{Pawlowski:2015PRL}
      N. Christiansen, M. Haas, J. M. Pawlowski, and N. Strodthoff,
          Phys. Rev. Lett. {\bf 115}, 112002 (2015).

\bibitem{Pawlowski:20156PRD}
      W. J. Fu, and J. M. Pawlowski,
          Phys. Rev. D {\bf 92}, 116006 (2015);
      W. J. Fu, and J. M. Pawlowski,
          Phys. Rev. D {\bf 93}, 091501 (R)  (2016).

\bibitem{Karsch:1996NPB}
      G. Boyd, J. Engels, F. Karsch, E. Laermann, C. Legeland, M. Lutgemeier, and B. Petersson,
          Nucl. Phys. B {\bf 469}, 419 (1996).

\bibitem{Karsch:2002LNP}
     F. Karsch,
       Lecture Notes in Phys. {\bf 585}, 209 (2002).

\bibitem{Forcrand:20023NPB}
     Ph. de Forcrand, O. Philipsen,
       Nucl. Phys.\ B {\bf 642}, 290 (2002);
       Nucl. Phys.\ B {\bf 673}, 170 (2003).
      Ph. de Forcrand, S. Kratochvila,
          Nucl. Phys. B Proc. Suppl. {\bf 153}, 62 (2006)

\bibitem{Fodor:20024JHEP}
     Z. Fodor, and S. D. Katz,
           arxiv: hep-ph 0908.3341, (2009).
%
     Z. Fodor, and S. D. Katz,
        Phys. Lett. B {\bf 534}, 87 (2002).
%
     Z. Fodor, and S. D. Katz,
        J. High Energy Phys. {\bf 03} (2002), 014.
%
        J. High Energy Phys. {\bf 04} (2004), 050.

\bibitem{DElia:2003PRD}
     M. D'Elia, and M. P. Lombardo,
       Phys. Rev. D {\bf 67}, 014505 (2003).


\bibitem{Bernard:2005PRD}
     C. Bernard, T. Burch, C. DeTar, J. Osborn, S. Gottlieb, E. B. Gregory, D. Toussaint, U. M. Heler,
     and R. Suger,
         Phys. Rev. D {\bf 71}, 034504 (2005).

\bibitem{Aoki:2006Nature}
     Y. Aoki, G. Endrodi, Z. Fodor, S. D. Katz, and K. K. Szabo,
       Nature {\bf 443}, 675 (2006);

\bibitem{Aoki:2006PLB}
     Y. Aoki, Z. Fodor, S. D. Katz, and K. K. Szabo,
      Phys. Lett. B {\bf 643}, 46 (2006).

\bibitem{Allton:2006PRD}
     C. R. Allton, M. Doring, S. Ejiri, S. J. Hands, O. Kaczmarek, F. Karsch, E. Laermann, and K. Redlich,
       Phys. Rev. D {\bf 71}, 054508 (2006).

\bibitem{Gavai:2005PRD}
     R. V. Gavai, and S. Gupta,
       Phys. Rev. D {\bf 71}, 114014 (2005);
       Phys.\ Rev.\ D {\bf 78}, 114503 (2008).


\bibitem{deForcrand:2014PRL}
     Ph. de Forcrand, and O. Philipsen,
       Phys. Rev. Lett. {\bf 105}, 152001 (2010);
%
     Ph. de Forcrand, J. Langelage, O. Philipsen, and W. Unger,
       Phys. Rev. Lett. {\bf 113}, 152002 (2014).

\bibitem{Endrodi:2011JHEP}
     G. Endrodi, Z. Fodor, S. D. Katz, and K. K. Szabo,
       JHEP {\bf 04} 001 (2011).

\bibitem{Kaczmarek:2011PRD}
     O. Kaczmarek, F. Karsch, E. Laermann, C. Miao, S. Mukherjee, P. Petreczky, C. Schmidt, W. Soeldner, and W. Unger,
       Phys. Rev. D {\bf 83}, 014504 (2011).

\bibitem{Li:2011PRD}
     A. Li, A.  Alexandru, and K. F. Liu,
        Phys. Rev. D {\bf 84}, 071503 (2011).

\bibitem{Karsch:2011PLB}
     F. Karsch,  B. J. Schaefer,  M.  Wagner, and J. Wambach,
         Phys. Lett. B {\bf 698}, 256 (2011).

\bibitem{Ding:2014PRL}
      T. Bhattacharya, M. I. Buchoff, N. H. Christ, {\it et al.},
         Phys. Rev. Lett. {\bf 113}, 082001 (2014).

\bibitem{Bazavov:2012PRD}
      A. Bazavov, T. Bhattacharya, M. Cheng, {\it et al.},
         Phys. Rev. D {\bf 85}, 054503 (2012).
         Phys. Rev. {\bf 86}, 034509 (2012).

\bibitem{Cea:2014PRD}
     P. Cea, L. Cosmai, and A. Papa,
       Phys. Rev. D {\bf 89}, 074512 (2014).

\bibitem{Gupta:2014PRD}
      S. Gupta, N. Karthik, and P. Majumdar,
        Phys. Rev. D {\bf 90}, 034001 (2014).

\bibitem{Cheng:2009PRD}
     M. Cheng, {\it et al.}, (for HotQCD Collaboration),
     Phys. Rev. D {\bf 79}, 074505 (2009).

\bibitem{Gavai:2011PLB}
    R. V. Gavai, and S. Gupta,
        Phys.\ Lett.\ B {\bf 696}, 459 (2011).

\bibitem{Bazavov:2012PRL}
     A. Bazavov, et. al. (for HotQCD Collaboration),
          Phys. Rev. Lett. {\bf 109}, 192302 (2012).

\bibitem{Borsanyi:2012JHEP}
   S. Borsanyi, Z. Fodor, S.D. Katz, S. Krieg, C. Ratti, and K. K. Szab\'{o},
    J. High Energy Phys. 01 (2012), 138.

\bibitem{Borsanyi:2013PRL}
     S. Borsanyi, Z. Fodor, S. D. Katz, S. Krieg, C. Ratti, and K. K. Szabo,
        Phys. Rev. Lett. {\bf 111}, 062005 (2013).

\bibitem{Borsanyi:2014PRL}
     S. Borsanyi, Z. Fodor, S. D. Katz, S. Krieg, C. Ratti, and K. K. Szabo,
         Phys. Rev. Lett. {\bf 113}, 052301 (2014).


\bibitem{Palhares:2010PRD}
     L. F. Palhares, and E. S. Fraga,
        Phys. Rev. D {\bf 82}, 125018 (2010).

\bibitem{Holl:1999PRC}
     A. Holl, P. Maris, and C. D. Roberts,
        Phys. Rev. C {\bf 59}, 1751 (1999).

\bibitem{Heiselberg:1993PRL}
     H. Heiselberg, C. J. Pethick, and E. F. Staubo,
       Phys. Rev. Lett. {\bf 70}, 1355 (1993).

\bibitem{Palhares:2011jd}
     L. F. Palhares and E. S. Fraga,
       PoS FACESQCD, 014 (2010).

\bibitem{Fraga:2015PRD}
   D. Kroff, and E. S. Fraga,
    Phys. Rev. D {\bf 91}, 025017 (2015).

\bibitem{Weber:2015PRC}
   S. M. de Carvalho, R. Negreiros, M. Orsaria, G. A. Contrera, F. Weber, and W. Spinella,
     Phys. Rev. C {\bf 92}, 035810 (2015).

\bibitem{Weber:2016PRC}
   I. F. Ranea-Sandoval, S. Han, M. G. Orsaria, G. A. Contrera, F. Weber, and M. G. Alford,
     Phys. Rev. C {\bf 93}, 045812 (2016).


\bibitem{Ke:2014PRD}
     W. Y. Ke, and Y. X. Liu,
       Phys. Rev. D {\bf 89}, 074041 (2014).

\bibitem{Randrup:2009PRC}
      J. Randrup,
         Phys. Rev. C {\bf 79}, 054911 (2009).


\bibitem{deForcrand:2004jt}
      P. de Forcrand, B. Lucini, and M. Vettorazzo,
          Nucl. Phys. Proc. Suppl. {\bf 140}, 647 (2005).

\bibitem{Pinto:2012PRC}
      M. B. Pinto, V. Koch, and J. Randrup,
         Phys. Rev. C {\bf 86}, 025203 (2012).


\bibitem{Mintz:2013PRD}
      B. W. Mintz, R. Stiele, R. O. Ramos, and J. Schaffner-Bielich,
         Phys. Rev. D {\bf 87}, 036004 (2013).

\bibitem{Song:2010PRC}
      J. Song, Z. T. Liang, Y. X. Liu, F. L. Shao, and Q. Wang,
        Phys. Rev. C {\bf 81}, 057901 (2010).

\bibitem{Peng:2010PRC}
   X. J. Wen, J. Y. Li, J. Q. Liang, and G. X. Peng,
     Phys. Rev. C {\bf 82}, 025809 (2010).

\bibitem{Lugones:2013PRC}
   G. Lugones, A. G. Grunfeld, and M. Al Ajmi,
     Phys. Rev. C {\bf 88}, 045803 (2013).


\bibitem{Pinto:2013PRC}
   A. F. Garcia, and M. B. Pinto,
     Phys. Rev. C {\bf 88} 025207 (2013).


\bibitem{Peng:2016PRD}
    C. J. Xia, G. X. Peng, E. G. Zhao, and S. G. Zhou,
      Phys. Rev. D {\bf 93}, 085025 (2016).


\bibitem{Blaizot:2001PRD}
   J.-P. Blaizot, E. Iancu, and A. Rebhan,
     Phys. Rev. Lett. {\bf 83}, 2906 (1999).
     Phys. Rev. D {\bf 63}, 065003 (2001).


\bibitem{Greco:2003PRC}
    V. Greco, C. M. Ko, and P. Levai,
      Phys. Rev. C {\bf 68}, 034904 (2003).

\bibitem{Yamazaki:2015NPA}
   K. Yamazaki, T. Matsui, and G. Baym,
     Nucl. Phys. A {\bf 933}, 245 (2015).

\bibitem{Roberts:DSEinitial}
    C. D. Roberts, and A. G. Williams,
         Prog. Part. Nucl. Phys. {\bf 33},  477 (1994);
%
    C. D. Roberts, and S. Schmidt,
        Prog. Part. Nucl. Phys. {\bf 45}, S1 (2000).
%

\bibitem{Alkofer:2001Infrared}
     R. Alkofer, and L. von Smekal,
         Phys. Rept. {\bf 353}, 281 (2001).

\bibitem{Roberts:20124Review}
     P. Maris, and C. D. Roberts,
        Int. J. Mod. Phys. E {\bf 12}, 297 (2003).
%
    A. Bashir, L. Chang, I. C. Cloet, B. El-Bennich, Y. X. Liu, C. D. Roberts, and P. C. Tandy,
        Commun. Theor. Phys. {\bf 58}, 79 (2012);
%
    I. C. Cloet, and C. D. Roberts,
       Prog. Part. Nucl. Phys. {\bf 77}, 1 (2014).


\bibitem{Wang:2012PRD}
     K. L. Wang, S. X. Qin, Y. X. Liu, L. Chang, C. D. Roberts, and S. M. Schmidt,
        Phys. Rev. D {\bf 86}, 114001 (2012).

\bibitem{Srednicki:1993PRL}
     M. Srednicki,
         Phys. Rev. Lett. {\bf 71}, 666 (1993).

\bibitem{Eisert:2010RMP}
     J. Eisert, M. Cramer, and M. B. Plenio,
        Rev. Mod. Phys. {\bf 82}, 277 (2010).

\bibitem{Haque:2013PRD}
      N.  Haque,  M. G. Mustafa, M. Strickland,
        Phys. Rev. D {\bf 87}, 105007 (2013).

\bibitem{Thoma:1998NPA}
     M. H. Thoma,
        Nucl. Phys. A {\bf 638}, 317C (1998).


\bibitem{Qin:2011PRC}
     S. X. Qin, L. Chang, Y. X. Liu, C. D. Roberts, and D. J. Wilson,
       Phys. Rev. C {\bf 84}, 042202(R) (2011).

\bibitem{Bowman:2004PRD}
     P. O. Bowman, U. M. Heller, D. B. Leinweber, M. B. Parappilly, and A. G. Williams,
       Phys. Rev. D {\bf 70}, 034509 (2004);
%
     P. O. Bowman, U. M. Heller, D. B. Leinweber, M. B. Parappilly, A. G. Williams, and J. B. Zhang,
        Phys. Rev. D. {\bf 71}, 054507 (2005).


\bibitem{Bogolubsky:2009PLB}
      I. Bogolubsky, E. Ilgenfritz, M. Muller-Preussker, and A. Sternbeck,
        Phys. Lett. B {\bf 676}, 69 (2009).

\bibitem{Boucaud:2010PRD}
     P. Boucaud, M. E. Gomez, J. P. Leroy, A. Le Yaouanc, J. Micheli, O. Pene, and J. Rodriguez-Quintero,
       Phys. Rev. D {\bf 82}, 054007 (2010).

\bibitem{Oliveira:2011JPG}
       O. Oliveira, and P. Bicudo,
          J. Phys. G {\bf 38}, 045003 (2011).

\bibitem{Cucchieri:2012PRD}
      A. Cucchieri, D. Dudal, T. Mendes, and N. Vandersickel,
          Phys. Rev. D {\bf 85}, 094513 (2012).

\bibitem{Aguilar:2012PRD}
     A. Aguilar, D. Binosi and J. Papavassiliou,
        Phys. Rev. D {\bf 86}, 014032 (2012).

\bibitem{Ayala:2012PRD}
     A. Ayala, A. Bashir, D. Binosi, M. Cristoforetti, and J. Rodriguez-Quintero,
         Phys. Rev. D {\bf 86}, 074512 (2012).

\bibitem{Dudal:2012PRD}
      D. Dudal, O. Oliveira, and J. Rodriguez-Quintero,
         Phys. Rev. D {\bf 86}, 105005 (2012).

\bibitem{Strauss:2012PRL}
      S. Strauss, C. S. Fischer, and C. Kellermann,
        Phys. Rev. Lett. {\bf 109}, 252001 (2012).

\bibitem{Weber:2012JPCS}
     A. Weber,
        J. Phys. Conf. Ser. {\bf 378}, 012042 (2012).

\bibitem{Zwanziger:2013PRD}
     D. Zwanziger,
        Phys. Rev. D {\bf 87}, 085039 (2013).

\bibitem{Blossier:2013PRD}
     B. Blossier, P. Boucaud, M. Brinet, F. De Soto, V. Morenas, O. Pene, K. Petrov,
     and J. Rodriguez-Quintero,
         Phys. Rev. D {\bf 87}, 074033 (2013).


\bibitem{Munczek:1995PRD}
      H. J. Munczek,
          Phys. Rev. D {\bf 52}, 4736 (1995).

\bibitem{Bender:1996PLB}
     A. Bender, C. D. Roberts, and  L. von Smekal,
        Phys. Lett. B {\bf 380}, 7 (1996).


\bibitem{Burden:1997PRC}
      C. J. Burden, L. Qian, C. D. Roberts, P. C. Tandy, and M. J. Thomson,
          Phys. Rev. C {\bf 55}, 2649 (1997).

\bibitem{Watson:2004FBS}
      P. Watson, W. Cassing, and P. C. Tandy,
         Few Body Syst. {\bf 35}, 129 (2004).

\bibitem{Maris:2007AIPCP}
     P. Maris,
       AIP Conf. Proc. {\bf 892}, 65 (2007).


\bibitem{Cloet:2007pi}
     I. C. Clo{\"e}t, A. Krassnigg, and C. D. Roberts,
 In {\it Proceedings of 11th International Conference on Meson-Nucleon Physics and the
  Structure of the Nucleon} \emph{(MENU 2007)},
  eds.\ H.~Machner and S.~Krewald, paper
  125.

\bibitem{Fischer:2009PRLb}
      C. S. Fischer, and R. Williams,
          Phys. Rev. Lett. {\bf 103}, 122001 (2009).

\bibitem{Krassnigg:2009PRD}
      A. Krassnigg,
         Phys. Rev. D {\bf 80}, 114010 (2009).

\bibitem{Krassnigg:2011PRD}
      A. Krassnigg, and M. Blank,
         Phys. Rev. D {\bf 83}, 096006 (2011).

\bibitem{Krassnigg:20115PRD}
      M. Blank, and A. Krassnigg,
         Phys. Rev. D {\bf 84}, 096014 (2011).
%
      T. Hilger, C. Popovici, M. Gomez-Rocha, and A. Krassnigg,
         Phys. Rev. D {\bf 91}, 034013 (2015).

\bibitem{Krassnigg:2015PRD}
      T. Hilger, M. Gomez-Rocha, and A. Krassnigg,
         Phys. Rev. D {\bf 91}, 114004 (2015).

\bibitem{Krassnigg:20156PRD}
      M. Gomez-Rocha, T. Hilger, and A. Krassnigg,
         Few-Body Systems {\bf 56}, 475 (2015);
      M. Gomez-Rocha, T. Hilger, and A. Krassnigg,
         Phys. Rev. D {\bf 92}, 054030 (2015);
%
      M. Gomez-Rocha, T. Hilger, and A. Krassnigg,
         Phys. Rev. D {\bf 93}, 074010 (2016).


\bibitem{Haymaker:1990vm}
     R. W. Haymaker,
        Riv. Nuovo Cim. {\bf 14N8}, 1 (1991).

\bibitem{kapusta1993finite}
      J. I. Kapusta,
        Finite-temperature field theory (Cambridge University Press, 1993).

\end{thebibliography}
\end{document}